\documentclass[11pt, preprint]{article}
\usepackage{color} \usepackage{cancel}
\usepackage{float} \usepackage{graphicx}
\usepackage{subfigure} \usepackage{caption}	
\usepackage{dcolumn}
\usepackage[toc,page]{appendix} \usepackage{authblk}	 		
\usepackage{indentfirst}	
\usepackage{authblk}	 		
\usepackage{multirow}     
\usepackage{hyperref}     
\usepackage{amsfonts, amssymb, amsmath}
\usepackage[left=2.5cm,right=2.5cm,top=2.9cm,bottom=2.9cm,a4paper]{geometry}
\usepackage[noadjust]{cite}
\usepackage{filecontents}

\usepackage{soul} 
\usepackage{xcolor}
\setstcolor{red}

\graphicspath{{./figures/}}
\hypersetup{ colorlinks   = true, 
urlcolor     = blue, 
linkcolor    = blue, 
citecolor    = red 	 
}
\newcommand{\bea}{\begin{eqnarray}} \newcommand{\ena}{\end{eqnarray}}
\newcommand{\beq}{\begin{equation}} \newcommand{\eeq}{\end{equation}}

\title{\Large\bf Reconstruction of the Scalar Field Potential in Inflationary Models with a Gauss-Bonnet term}
\author[1]{Seoktae Koh\thanks{kundol.koh@jejunu.ac.kr}}
\author[2,3]{Bum-Hoon Lee\thanks{bhl@sogang.ac.kr}}
\author[3]{Gansukh Tumurtushaa\thanks{gansuh@sogang.ac.kr}}
\affil[1]{\small \it Department of Science Education, Jeju National University, Jeju, 63243, Korea }
\affil[2]{\small \it Asian Pacific Center for Theoretical Physics, Pohang 37673, Korea}
\affil[3]{\small \it Center for Quantum Spacetime, Sogang University, Seoul 121-742, Korea \\
Department of Physics, Sogang University, Seoul 121-742, Korea}

\begin{document}

\maketitle
\begin{abstract}
We study inflationary models with a Gauss-Bonnet term to reconstruct the scalar field potentials and the Gauss-Bonnet coupling functions  from the observable quantities. Using the observationally favored relations for both  $n_s$ and $r$, we derive the expressions for both the scalar field potentials and the coupling functions. The implication of the blue-tilted spectrum, $n_t>0$, of the primordial tensor fluctuations is discussed for the reconstructed configurations of the scalar field potential and the Gauss-Bonnet coupling.
\end{abstract}
\section{Introduction}

Current stage of the cosmological observations, particularly of the Cosmic Microwave Background (CMB) anisotropies by the WMAP \cite{Hinshaw:2012aka} and Planck collaborations \cite{Ade:2013zuv}\cite{Ade:2015xua}, have the prospect of providing precise constraints on the models for the origin of the large-scale structure, amongst which inflation~\cite{Guth:1980zm} is currently the leading candidate. The inflationary scenario, there exist hundreds of different inflationary scenarios in the market at present~\cite{Martin:2013tda}, remains successful to explain the current observations.

In spite of the success of inflation, there have been several unresolved problems, {\it for example} the initial singularity problem and trans-Planckian problem.  Especially, if we consider  the Planck scale, it is widely believed that quantum gravity would play an important role. However, because we have any complete theory of quantum gravity yet, we would consider Einstein gravity with some modifications as the effective theory of  ultimate quantum gravity. One of the corrections is to consider the Gauss-Bonnet term which appears naturally in low energy effective theory of string theory and in renormalising  the stress tensor in curved spacetime.
We have considered the Gauss-Bonnet term to study the importance of theories beyond standard single field slow-roll inflation models in our previous paper~\cite{Koh:2014bka}. If a particular form of the potential is given, in general, one computes the observable quantities. As a result of the work, we computed the observable quantities for the specific choices of the potentials and the coupling functions and provided constraints on those quantities in light of observational data~\cite{Ade:2013zuv}.

If a particular set of observations with some accuracy is given, one can attempt the task to reconstruct the inflaton potential from the observable quantities~\cite{Copeland:1993jj}\cite{Chiba:2015zpa}. Therefore, in our current work, we are interested in the inverse problem of reconstructing the inflaton potentials and the Gauss-Bonnet coupling functions from the observable quantities. Following the approach used in Ref.~\cite{Chiba:2015zpa}, we extend the study to the inflationary models with the Gauss-Bonnet term that is non-minimally coupled to a dynamical scalar field.

One interesting feature in the inflation model with the Gauss-Bonnet term is that the consistency relation $r=-8n_t$ of the standard inflation model is violated. In the conventional inflation model with the scalar field minimally coupled to gravity, the Hubble rate monotonically decreases ($\dot{H}<0$), such that $\epsilon>0$. Hence, one can conclude that the spectral index of the primordial tensor fluctuation is always negative, $n_t= -2\epsilon$. Therefore, the spectrum of the tensor modes is red-tilted. Although the present observations cannot determine the tilt of the tensor spectral index, from the perspective of theoretical interpretations, it is interesting to investigate the blue spectrum of the tensor modes in the framework of inflationary cosmology~\cite{Cai:2014uka}.

This paper is organized as follows. In Section~\ref{section2}, we briefly review the main findings of our previous work and describe the procedure for constructing the inflaton potential. We consider example models in Section~\ref{section3} to reconstruct the inflaton potential as well as the Gauss-Bonnet coupling function. Section~\ref{section4} illustrates what can be learned for the reconstructed potential from the observable quantities. In our model, the blue spectrum for the tensor fluctuations is obtained and that we will discuss the details in the section. We conclude with a discussion in Section~\ref{section5}. The units of $\kappa^2=8\pi G = M_\text{pl}^{-2}$ is used throughout this paper.

\section{Setup}\label{section2}

The action that we consider is composed of the Einstein-Hilbert term and the canonical scalar field which couples non-minimally to the Gauss-Bonnet term through the coupling function $\xi(\phi)$,
\begin{align}
S = &\int d^4x\sqrt{-g}
\left[\frac{1}{2\kappa^2} R - \frac{1}{2}g^{\mu\nu}
\partial_{\mu}\phi \partial_{\nu} \phi
 - V(\phi)-\frac12\xi(\phi) R_{\rm GB}^2\right],
\label{action}
\end{align}
where $R^{2}_{\rm GB} = R_{\mu\nu\rho\sigma} R^{\mu\nu\rho\sigma}- 4 R_{\mu\nu} R^{\mu\nu} + R^2$ is the Gauss-Bonnet term. The Gauss-Bonnet coupling $\xi(\phi)$ is required to be a function of the scalar field in order to give nontrivial effects on the background dynamics. In a Friedmann-Robertson-Walker(FRW) universe with a scale factor $a$ and with an arbitrary constant curvature $K$,
\begin{align}
ds^2 = - dt^2 + a^2\left(\frac{d r^2}{1-K r^2}+r^2 d\Omega^2\right),
\end{align}
the background dynamics of this system yields the Einstein and the field equations
\bea \label{beq2a}
&& H^2 = \frac{\kappa^2}{3} \left[\frac{1}{2}\dot{\phi}^2 + V -\frac{3K}{\kappa^2 a^2}+ 12\dot{\xi}H\left(H^2+\frac{K}{a^2}\right) \right],\\
\label{beq3a}
&& \dot{H} = -\frac{\kappa^2}{2}\left[\dot{\phi}^2-\frac{2 K}{\kappa^2 a^2} -4\ddot{\xi}\left(H^2+\frac{K}{a^2}\right)-4\dot{\xi}H\left(2\dot{H}-H^2-\frac{3K}{a^2}\right) \right], \\
&& \ddot{\phi} + 3 H \dot{\phi} + V_{\phi} +12 \xi_{\phi} \left(H^2+\frac{K}{a^2}\right) \left(\dot{H}+H^2\right) = 0, \label{beq4a}
\ena
where a dot represents a derivative with respect to the cosmic time $t$,
$H \equiv \dot{a}/a$ denotes the Hubble parameter, $V_{\phi} = \partial V/\partial \phi$, and $\xi_{\phi} = \partial \xi/\partial \phi$. Since $\xi$ is a function of $\phi$,
$\dot{\xi}$ implies $\dot{\xi} = \xi_{\phi} \dot{\phi}$. If $\xi$ is a constant,
the background dynamics would not be influenced by the Gauss-Bonnet term because it is known that the Gauss-Bonnet in four dimensional spacetime is a topological term.

In this work, we consider the case in which the scalar field slowly rolls down to the minimum of the potential and the Gauss-Bonnet term is assumed to be a small correction to gravity. Hence, the following inequality must be satisfied~\cite{Koh:2014bka};
\begin{align}
\dot{\phi}^2/2  \ll V\,, \quad \ddot{\phi} \ll 3H\dot{\phi}\,,
\quad 4\dot{\xi}H \ll 1\,, \quad \text{and} \quad\ddot{\xi} \ll \dot{\xi} H\,.
\label{sra1}
\end{align}
We define the following slow-roll parameters to reflect the slow-roll approximations above
\bea\label{eq:sl_param}
\epsilon\equiv-\frac{\dot{H}}{H^2}\,,\,\,\, \eta\equiv\frac{\ddot{H}}{H\dot{H}}\,,\,\,\,\delta_1\equiv4\kappa^2\dot{\xi}H\,,\,\,\,\delta_2\equiv\frac{\ddot{\xi}}{\dot{\xi}H}\,.
\ena
Under Eq.~(\ref{sra1}), the background equations, Eqs.~(\ref{beq2a})--(\ref{beq4a}), become for $K=0$
\bea
\label{seq1}
&& H^2 \simeq \frac{\kappa^2}{3}  V\,, \quad \\
&& \dot{H} \simeq -\frac{\kappa^2}{2}( \dot{\phi}^2 + 4\dot{\xi}H^3)\,,\label{seq3}\quad\\
&& 3H \dot{\phi} + V_{\phi} + 12\xi_{\phi}H^4 \simeq 0\,.
\label{seq2}
\ena
We rewrite Eq.~(\ref{eq:sl_param}) in terms of the potential and the Gauss-Bonnet coupling function as
\bea\label{eq:srpote}
\epsilon &=& \frac{1}{2\kappa^2} \frac{V_{\phi}}{V}Q\,,
\\
\eta &=& -\frac{1}{\kappa^2}\left(\frac{V_{\phi\phi}}{V_{\phi}}Q + Q_{\phi}\right)\,,
\\
\delta_1 &=& -\frac{4\kappa^2}{3} \xi_{\phi} V Q,\label{eq:sl_d1phi}
\\
\delta_2 &=& -\frac{1}{\kappa^2}\left(\frac{\xi_{\phi\phi}}{\xi_{\phi}}Q+\frac{1}{2}\frac{V_{\phi}}{V}Q+Q_{\phi}\right)\,,\label{eq:sl_d2phi}
\ena
where
\bea\label{eq:QofN}
Q &\equiv& \frac{V_{\phi}}{V} + \frac{4}{3}\kappa^4 \xi_{\phi} V\,.
\ena
Another key parameter in an inflationary scenario is the $e$-folding number, $N$, that measures the amount of inflationary expansion from a particular time $t$ until the end of inflation $t_e$
\bea
\label{eq:ne}
N =\int^{t_e}_{t} Hdt \simeq \int^{\phi}_{\phi_e} \frac{\kappa^2}{Q} d\phi,
\ena
where $\phi_e=\phi(t_e)$ is the field value at the end of inflation. To give standard reheating process, $N\simeq 50\sim 60$ is assumed at the horizon crossing time, $k=a H$ where $k$ is the comoving scale.

If the potential and the Gauss-Bonnet coupling function are given, the observable quantities can be easily obtained~\cite{Koh:2014bka} up to leading order in terms of the slow-roll parameters as
\bea\label{eq:nsob}
n_s -1 &\approx& -2\epsilon
- \frac{2\epsilon (2\epsilon +\eta) -\delta_1 (\delta_2 -\epsilon)}{2\epsilon - \delta_1}\,,
\\
r &\approx&  8 (2\epsilon -\delta_1)\,,\label{eq:rob}
\\
n_t &\approx& -2\epsilon\,.\label{eq:ntob}
\ena
After computing Eqs.~(\ref{eq:nsob})--(\ref{eq:ntob}) for a given potential and the Gauss-Bonnet coupling function, the rest is to check its consistency with the observational data.

However, in this work, we are interested in an inverse problem of reconstructing the inflaton potential $V(\phi)$ and the Gauss-Bonnet coupling function $\xi(\phi)$ from the observational data~\cite{Ade:2013zuv, Ade:2015xua} using Eqs.~(\ref{eq:nsob})--(\ref{eq:ntob}).
To reconstruct $V(\phi)$ and $\xi(\phi)$, we use $n_s$ and $r$ that are functions of $N$. Therefore, first, we construct them in terms of $N$, then write $N$ as a function of $\phi$ by using Eq.~(\ref{eq:ne}).

Since the observable quantities can be expressed as the functions of $N$~\cite{Hinshaw:2012aka, Ade:2013zuv, Ade:2015xua}, it is convenient to work with the slow-roll parameters as the functions of $N$ and we obtain
\bea\label{srpote}
\epsilon &=&\frac{1}{2}\frac{V_{N}}{V}\,,
\\
\eta &=&-\frac{V_{NN}}{V_N}=-2\epsilon-\frac{d\ln \epsilon}{dN}\,,
\\
\delta_1 &=& -\frac{4}{3}\kappa^4\xi_{N}V\,,\label{eq:sl_d1}
\\
\delta_2 &=&-\frac{\xi_{NN}}{\xi_N}-\frac{1}{2}\frac{V_N}{V}=\epsilon-\frac{d\ln\delta_1}{dN}\label{delta2}\,.
\ena
By using Eqs. (\ref{srpote})--(\ref{delta2}), we rewrite Eqs.~(\ref{eq:nsob})--(\ref{eq:ntob}) as
\bea\label{specind}
n_s(N) -1 &=& \left[\ln\left(\frac{V_N}{V^2}+\frac{4}{3}\kappa^4\xi_N\right)\right]_{,N}\,,
\\
r(N) &=&8\left(\frac{V_{N}}{V}+\frac{4}{3}\kappa^4\xi_{N}V\right)=8Q^{(N)}\label{ttsr}\,,
\\
n_t(N) &=&-\frac{V_{N}}{V}\,,
\ena
where $[\dots]_{,N}$ represents a derivative respect to the $e$-folding number $N$. 
We obtain the scalar field potential  in terms of $n_s$ and $r$ from Eqs. (\ref{specind}) and (\ref{ttsr}),
\bea\label{eq:potofN}
V(N)=\frac{1}{8c_1}r(N)e^{-\int[n_s(N)-1]dN}\,,
\ena
and then we find $\xi(N)$ by substituting Eq.~(\ref{eq:potofN}) into Eq.~(\ref{ttsr}),
\bea\label{eq:xi}
\xi(N)=\frac{3}{4\kappa^4} \left[\frac{1}{V(N)}+\int\frac{r(N)}{8V(N)}dN + c_2\right]\,,
\ena
where $c_1$ and $c_2$ are the integration constants.
Therefore, for the given relations of $n_s-1$ and $r$, one can construct the scalar field potential and the Gauss-Bonnet coupling functions. Using Eq.~(\ref{eq:ne}) together with Eq.~(\ref{ttsr}), one also can find the relation between the number of e-folding $N$ and the scalar field $\phi$ as,
\bea\label{eq:met2nofphi}
\int_{\phi_{e}}^\phi d\phi=\int\sqrt{\frac{r(N)}{8\kappa^2}}dN \,.
\ena

\section{Example models}\label{section3}
There are hundreds of inflation models in the market~\cite{Martin:2013tda} that show good fit with the observational data, hence it is hard to figure out a unique inflation model even when the model parameters accurately fit with the data. Therefore, in this section, we reconstruct the inflaton potentials as well as the Gauss-Bonnet coupling functions by using the general relations for both $n_s$ and $r$ that are in good agreement with the latest Planck data~\cite{Ade:2013zuv, Ade:2015xua}. As the input, we consider
\bea\label{eq:rel1}
n_s-1&=&-\frac{\beta}{N+\alpha}\,,\\
r&=&\frac{q}{N^p+\gamma N+\alpha}\,,\label{eq:rel2}
\ena
where $\beta$, $\gamma$, $p$ and $q$ parameters are arbitrary integers while $\alpha$ is also an arbitrary constant but not necessary to be an integer. These model parameters can be chosen such a way that the  relations in Eqs.~(\ref{eq:rel1})--(\ref{eq:rel2}) to be consistent with the  observational data~\cite{Ade:2013zuv, Ade:2015xua}.

Previously the authors of Ref.~\cite{Chiba:2015zpa} have studied the inverse problem of reconstructing the  inflaton potential from the   spectral index for a model without the Gauss-Bonnet term. They used the same relation as Eq.~(\ref{eq:rel1}), but without $\alpha$, to construct the potential. As the result, the authors obtained the tensor-to-scalar ratio, $r$, with the similar form as Eq.~(\ref{eq:rel2}) without $\alpha$. In our case, if the Gauss-Bonnet coupling in Eq.~(\ref{action}) is constant or zero, the consequent result must converge to that of Ref.~\cite{Chiba:2015zpa} because it is known that the Gauss-Bonnet term is topological in four dimensions. Therefore, in principle,  the scalar field potential that obtained in Ref.~\cite{Chiba:2015zpa} can be reproduced in our model.

In the following two subsections, we work with the specific models to construct the scalar field potentials and the Gauss-Bonnet coupling functions from Eqs.~(\ref{eq:rel1})--(\ref{eq:rel2}). In Section~\ref{subsec:gamma1}, we aim to test our method by reproducing the scalar field potential that obtained in Ref.~\cite{Chiba:2015zpa}. Therefore, without loss of generality, we set $\gamma=1$. Then, in Section~\ref{subsec:gamma0}, we extend our studies to other examples.

\subsection{Model with $\gamma=1$}\label{subsec:gamma1}

To be consistent with~\cite{Chiba:2015zpa}, we set $\beta=p=2$ and $q=8$ such that Eqs.~(\ref{eq:rel1})--(\ref{eq:rel2}) become
\bea\label{eq:gam1ns}
n_s-1&=&-\frac{2}{N+\alpha}\,,\\
r&=&\frac{8}{N^2+N+\alpha}\,.\label{eq:gam1r}
\ena
Then we  obtain  from Eqs.~(\ref{eq:xi}) and (\ref{eq:potofN}), 
\bea
V(N)&=&\frac{(N+\alpha)^2}{c_1 \left(N^2+N+\alpha\right)}\,,\label{eq:gam1VN}\\
\xi(N)&=&-\frac{3}{4 \kappa ^4} \left[-\frac{N^2}{(N+\alpha)^2}c_1+c_2\right]\,.\label{eq:gam1xiN}
\ena
Because $N$ can be solved in terms of $\phi$ from  Eq.~(\ref{eq:met2nofphi}),
\bea\label{eq:gam1N}
N=\frac{1}{4} \left[(1-4 \alpha ) e^{-\kappa  (\phi-C)}+e^{\kappa(\phi-C)}-2\right]\,,
\ena
where $C$ is an integration constant which is responsible for the shift of $\phi$, we rewrite both the potential and the coupling functions in terms of scalar field, $\phi$, as
\bea\label{eq:potgam1}
V(\phi)&=&\frac{\left(e^{\kappa  (\phi -C)}-1\right)^2 \left(4 \alpha +e^{\kappa  (\phi -C)}-1\right)^2}{c_1 \left(4 \alpha +e^{2 \kappa  (\phi -C)}-1\right)^2}\,,\\
\xi(\phi)&=&\frac{3}{4 \kappa ^4}\left[\frac{\left(1-2 e^{\kappa(C-\phi )}-(4 \alpha -1) e^{2 \kappa  (C-\phi )}\right)^2}{\left(1- e^{-\kappa  (\phi -C)}\right)^2 \left(1+(4 \alpha -1) e^{-\kappa  (\phi -C)}\right)^2}c_1-c_2\right]\,.\label{eq:xigam1}
\ena
In $\alpha\rightarrow0$ limit, the leading order contribution gives
\bea\label{eq:potalpha0}
V(\phi)&=&c_1\tanh ^2\left(\frac{1}{2} \kappa  (\phi-C )\right)\,,\\ 
\xi(\phi)&=&\frac{3}{4\kappa^4}\left(c_1-c_2\right)\,.\label{eq:xialpha0}
\ena
The Gauss-Bonnet coupling function in Eq.~(\ref{eq:xialpha0}) becomes zero if $c_1= c_2$ or constant otherwise. In either cases, the Gauss-Bonnet term does not give any effects on the background evolution in four dimensions. Therefore, the background evolution in our model reduces to Einstein gravity, and the Gauss-Bonnet correction term does not play any role on the dynamics in $\alpha\rightarrow0$ limit. Eq.~(\ref{eq:potalpha0}) shows that our result reproduces the potentials obtained in~\cite{Chiba:2015zpa}\cite{Kallosh:2013hoa}.

\subsection{Model with $\gamma=0$}\label{subsec:gamma0}

In this section, we consider $\gamma=0$ case to obtain the scalar field potential as well as the Gauss-Bonnet coupling function. For $\gamma=0$, Eqs.~(\ref{eq:rel1})--(\ref{eq:rel2}) become
\bea\label{eq:specind}
n_s(N)-1&=&-\frac{\beta}{N+\alpha}\,,\\
r(N)&=&\frac{q}{N^p+\alpha}\,.\label{eq:ttsr}
\ena
We obtain after substituting Eqs.~(\ref{eq:specind})--(\ref{eq:ttsr}) into Eq.~(\ref{eq:xi})
\bea\label{eq:xinew}
\xi(N)=\frac{3}{4\kappa^4}\left[\left(\frac{8}{q}\frac{N^p+\alpha}{(N+\alpha)^{\beta}} +\frac{(N+\alpha)^{1-\beta}}{1-\beta}\right)c_1+c_2\right]\,,
\ena
where $\beta\neq1$ is assumed.~\footnote{We do not consider a case with $\beta=1$, in this paper, due to our interest. \textcolor{black}{If $\beta=1$, from Eq.~(\ref{eq:specind}), the number of $e$-folds should approximately be $N \sim 30-\alpha$ in order to be consistent with the observational value of $n_s\sim0.9655\pm0.0062$~\cite{ Ade:2015xua}. On the other hand, we need $N\simeq50\sim60$ for inflation, therefore $\alpha$ must take negative value between $-30\leq\alpha\leq-20$ which later conflicts with Eq.~(\ref{eq:phiofN}) where $\alpha>0$ is necessary. For $\beta=2$ case, however, we have no such contradictions and everything is fine.}}
The scalar-field potential can also be obtained with the help of Eq.~(\ref{eq:potofN}) as
\bea\label{eq:potofNnew}
V(N)=\frac{q}{8c_1}\frac{(N+\alpha)^{\beta }}{N^p+\alpha}\,.
\ena
Eq.~(\ref{eq:met2nofphi}) gives
\bea\label{eq:phiofN}
\phi-\phi_e&=&N\sqrt{\frac{q}{8\kappa^2\alpha }} \, _2F_1\left(\frac{1}{2},\frac{1}{p};1+\frac{1}{p};-\frac{N^p}{\alpha }\right)\,,
\ena
where $\alpha>0$ is necessary since the  observations~\cite{Ade:2013zuv, Ade:2015xua} favor positive $q$, and this equation is to be solved as $N(\phi)$ for given $p$. It is often assumed for large field inflation that the field value at the end of inflation is negligible compared to that of the beginning of inflation, $\phi_e\ll\phi$. Therefore, from now and through out the rest of this paper, we will ignore $\phi_e$ with having in mind that we are dealing with the large field inflation model. In the following two subsections, we will consider $p=1$ and $p=2$ cases, respectively, for simplicity.

\subsubsection{$p=1$ case}
\begin{figure}[h!]
\centering
\subfigure[]
{\includegraphics[width=0.495\textwidth]{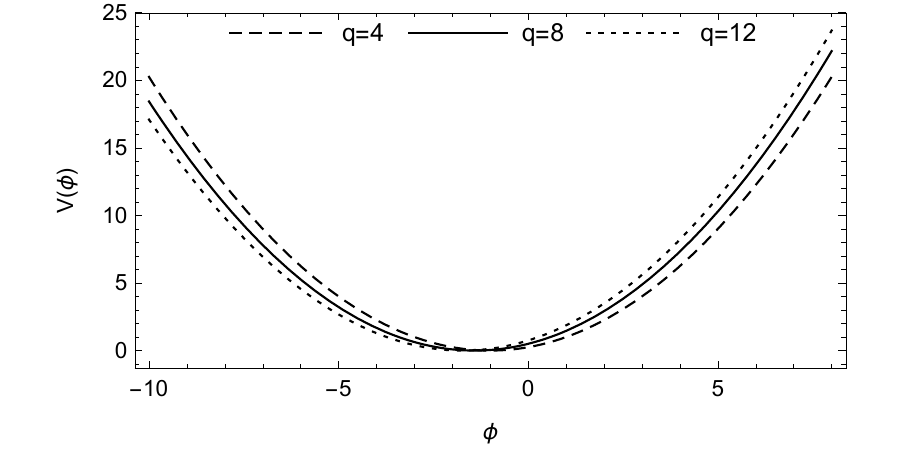}}
\subfigure[]
{\includegraphics[width=0.495\textwidth]{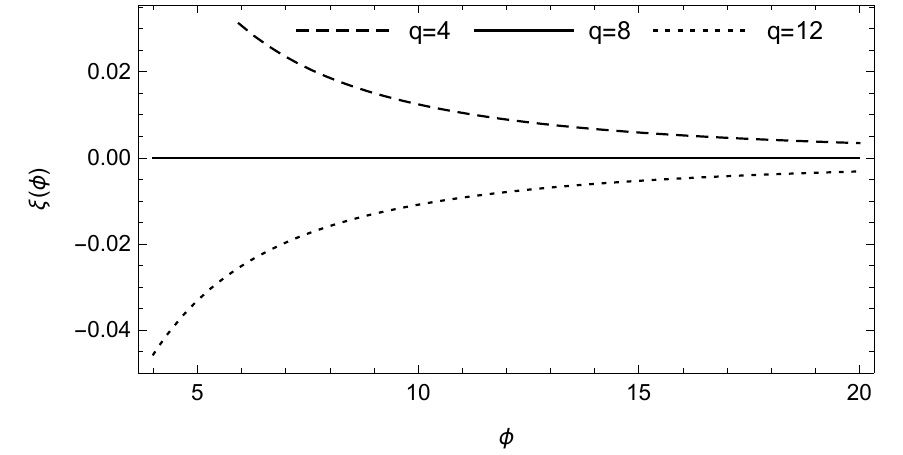}}\\
\caption{Numerical plot of Eq.~(\ref{eq:potofphi}) and Eq.~(\ref{eq:xiofphi}) with $c_1=1$, $c_2=0$, $\kappa^2=1$ and $n=2$. }
\label{fig:fig1}
\end{figure}

If $p=1$ in Eq.~(\ref{eq:ttsr}), the corresponding $n_s$ and $r$ relations look same as those obtained in Ref.~\cite{Guo:2010jr}. Therefore, if our reconstruction method is right, one may expect to see the power-law potential and the  inverse power-law coupling functions as a result in the end of this section.
When $p=1$, we obtain from Eq.~(\ref{eq:phiofN})
\bea\label{eq:Nofphi}
N=\left(\frac{2\kappa^2}{q}\phi^2+\sqrt{\frac{8\alpha}{q}}\kappa\phi\right),
\ena
where $q\neq0$ and Eqs.~(\ref{eq:xinew})--(\ref{eq:potofNnew}) give
\bea\label{eq:phiofab}
V(\phi)&=&\frac{q}{8c_1} \left(\alpha+\frac{2}{q}\kappa^2\phi^2+\sqrt{\frac{8\alpha}{q}}\kappa\phi\right)^{\beta-1}\,,\\
\xi(\phi)&=&\frac{3}{4\kappa^4}\left[\frac{q+8(1-\beta)}{q(1-\beta)} \left(\alpha+\frac{2}{q}\kappa^2\phi^2+\sqrt{\frac{8\alpha}{q}}\kappa\phi\right)^{1-\beta}c_1+c_2\right]\,.\label{eq:xiofab}
\ena
To be more consistent with  Ref.~\cite{Guo:2010jr}, it is worth to express $\alpha$ and $\beta$ in terms of the  new parameter, $n$, as follows
\bea
\alpha=\frac{n}{4}\,, \quad \beta=\frac{n+2}{2}\,.
\ena
Eqs.~(\ref{eq:phiofab})--(\ref{eq:xiofab}) can be rewritten as
\bea\label{eq:potofphi}
V(\phi)&=&\frac{q}{8c_1} \left(\frac{n}{4}+\frac{2}{q}\kappa^2\phi^2+\sqrt{\frac{2n}{q}}\kappa\phi\right)^{\frac{n}{2}}\,,\\
\xi(\phi)&=&\frac{3}{4\kappa^4}\left[\frac{8n-2q}{nq} \left(\frac{n}{4}+\frac{2}{q}\kappa^2\phi^2+\sqrt{\frac{2n}{q}}\kappa\phi\right)^{-\frac{n}{2}}c_1+c_2\right]\,,\label{eq:xiofphi}
\ena
where if $c_2=0$, the Gauss-Bonnet coupling function holds inverse relation to power-law potential, $\xi(\phi)\sim1/V(\phi)$.
In Fig.~\ref{fig:fig1}, we plot Eqs.~(\ref{eq:potofphi})--(\ref{eq:xiofphi}) for $n=2$.
As is seen in Fig.~\ref{fig:fig1} and in Eq.~(\ref{eq:xiofphi}), $\xi(\phi)=0$ for every $q=4n$ when $c_2=0$ or $\xi(\phi)=\text{const.}$ when $c_2\neq0$. In either case, the background evolution would be described by Einstein gravity with a dynamical scalar field. The potential takes its minimum value at $\phi_{\text{min}}=0$ for $\alpha=0$, however, the minimum shifts as
\bea\label{eq:potmin}
\phi_{\text{min}}=-\sqrt{\frac{nq}{8\kappa^2}}\,,
\ena
for $\alpha\neq0$ depending on both $q$ and $n$ values.

\subsubsection{$p=2$ case}
\begin{figure}[H]
\centering
\subfigure[]
{\includegraphics[width=0.495\textwidth]{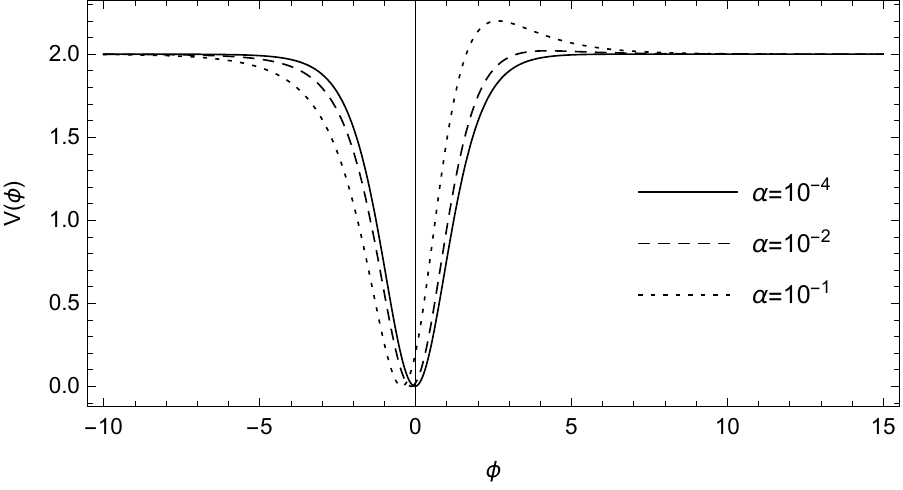}\label{fig:figp2a}}
\subfigure[]
{\includegraphics[width=0.495\textwidth]{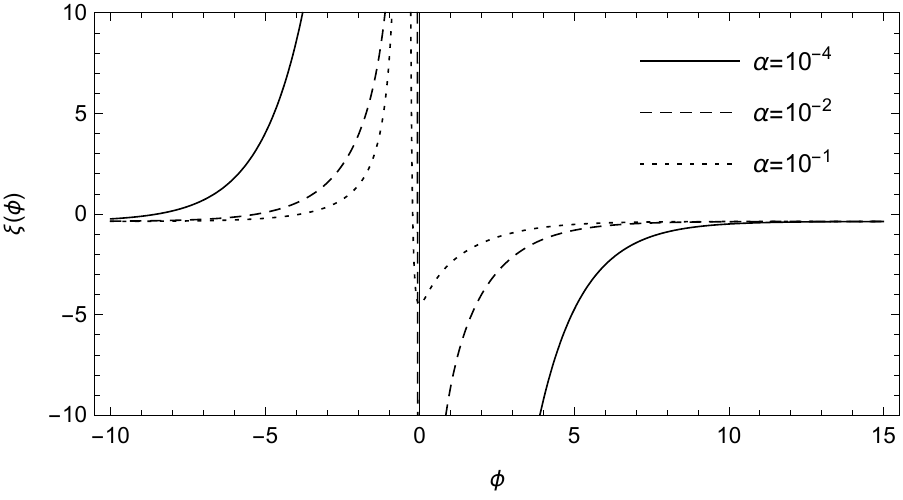}\label{fig:figp2b}}\\
\subfigure[]
{\includegraphics[width=0.495\textwidth]{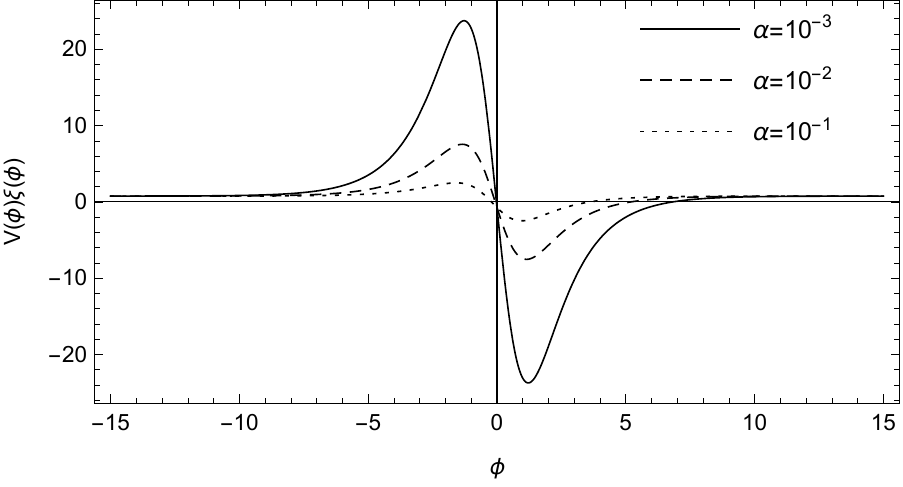}\label{fig:figp2c}}\\
\caption{Numerical plots of Eqs.~(\ref{eq:potofphinew})--(\ref{eq:xiofphinew}) with $c_1=1$, $c_2=0$, $\kappa^2=1$, $\beta=2$ and $q$=16. The bump shown in Fig.~\ref{fig:figp2a} increases as $\alpha$ increases or decreases as $\alpha$ decreases. }
\label{fig:figp2}
\end{figure}

\begin{figure}[H]
\centering
\includegraphics[width=0.6\textwidth]{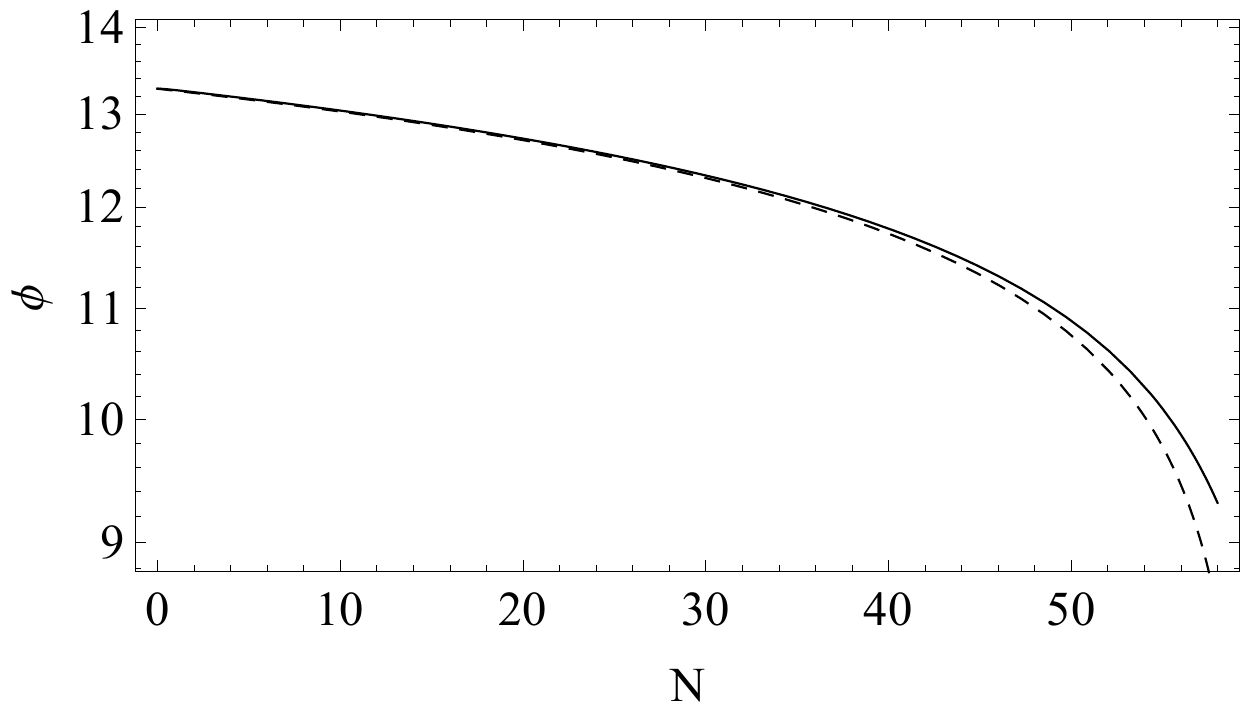}
\caption{Plot of the  numerical solution (solid) of Eqs.~(\ref{beq2a})--(\ref{beq4a}) when $K=0$ and the  slow-roll solution (dashed) obtained Eq.~(\ref{seq3}). We set parameters as; $c_1=1$, $c_2=0$, $\kappa^2=1$, $\alpha=10^{-4}$, $\beta=2$, $q=16$, $\phi_0=13.28$, $C=8.4\times10^{3}$ and $\dot{\phi}_0=0$. }\label{fig:compared}
\end{figure}

We obtain from Eq.~(\ref{eq:phiofN}) when $p=2$
\bea\label{eq:Nofphip2}
N=\sqrt{\alpha}\sinh\left(\sqrt{\frac{8}{q}}\kappa\phi\right)\,,
\ena
where $\alpha>0$ and $q>0$ are assumed. By substituting Eq.~(\ref{eq:Nofphip2}) into Eqs.~(\ref{eq:xinew})--(\ref{eq:potofNnew}), we obtain
\bea\label{eq:potofphinew}
V(\phi)&=&\frac{q}{8c_1\alpha}\text{sech}^2\left(\sqrt{\frac{8}{q}}\kappa\phi\right) \left[\alpha+\sqrt{\alpha}\sinh\left(\sqrt{\frac{8}{q}}\kappa\phi\right)\right]^{\beta }\,,\\
\xi(\phi)&=&\frac{3}{4\kappa^4} \left[\frac{q\left(\alpha+\sqrt{\alpha}\sinh\left(\sqrt{\frac{8}{q}}\kappa\phi\right)\right) +8(1-\beta)\alpha\cosh^2\left(\sqrt{\frac{8}{q}}\kappa\phi\right)} {q(1-\beta)\left(\alpha+\sqrt{\alpha}\sinh\left(\sqrt{\frac{8}{q}}\kappa\phi\right)\right)^{\beta}}c_1+c_2\right]\,,\label{eq:xiofphinew}
\ena
where $\beta\neq1$. In Fig.~\ref{fig:figp2}, we plot Eqs.~(\ref{eq:potofphinew})--(\ref{eq:xiofphinew}) with $\beta=2$. Previously, in $p=1$ case, we were able to see that the  potential and the  Gauss-Bonnet coupling functions hold inverse relation to each other when $c_2=0$. Unfortunately such relation seems not to be hold in $p=2$ case even when $c_2\neq0$ as shown in Fig.~\ref{fig:figp2c}.
In $\alpha\rightarrow 0$ limit, Eqs.~(\ref{eq:potofphinew})--(\ref{eq:xiofphinew}) reduce to
\bea\label{eq:tanhpot}
V(\phi)&\sim& \tanh^2\left(\sqrt{\frac{8}{q}}\kappa\phi\right)\,,\\
\xi(\phi)&\sim&-\frac{3 c_1}{4 \sqrt{\alpha } \kappa ^4}\text{csch}\left(\sqrt{\frac{8\kappa^2}{q}} \phi \right)\,.
\ena
Fig.~\ref{fig:figp2a} shows that the general shape of the reconstructed potential is similar to that of a ``T-model" studied in Ref.~\cite{Kallosh:2013hoa} with a small bump on the side which eventually disappears as $\alpha$ goes to zero and vice versa. The existence of such bump in the potential may play an important role for possible blue-tilt of the primordial tensor fluctuations as it is discussed in Ref.~\cite{Satoh:2010ep}. Before going into the details of the blue spectrum for the tensor modes in the next section, let us obtain the  slow-roll solution to the background equations of motion. Recalling approximate Eqs.~(\ref{seq1})--(\ref{seq2}) together with Eq.~(\ref{eq:potofphinew}) and Eq.~(\ref{eq:xiofphinew}) with $\beta=2$, we obtain the  following slow-roll solution for the  scalar field in terms of $N$,
\bea\label{seq3}
\phi(N)&=&-\sqrt{\frac{q}{8\kappa^2}}\text{arcsinh}\left(\frac{N}{\sqrt{\alpha}}-\sqrt{\frac{8\kappa^2}{q}}C\right),
\ena
where $C$ is an arbitrary constant. In Fig.~\ref{fig:compared}, we compare the numerical solution of Eqs.~(\ref{beq2a})--(\ref{beq4a}) with the  slow-roll solution in Eq.~(\ref{seq3}). It has been shown in Fig.~\ref{fig:compared} that the  slow-roll solution fits well with the  exact and numerical solution during inflation period.
\section{The blue-tilted spectrum of the tensor modes in $\gamma=0$ model with $p=2$}\label{section4}
According to Ref.~\cite{Satoh:2010ep}, an interesting feature of our model is that the spectrum of the primordial tensor fluctuations can be blue-tilted if the potential and the Gauss-Bonnet coupling functions take the form given in Eqs.~(\ref{eq:potofphinew})--(\ref{eq:xiofphinew}) when $\beta=2$. The blue-tilt of the tensor fluctuations is impossible to be achieved for the  conventional inflation models, those considered in Ref.~\cite{Ade:2013zuv, Ade:2015xua}. The Hubble rate $H$ monotonically decreases during slow-roll inflation for these conventional models of inflation, $\dot{H}<0$, hence it is implied that $\epsilon>0$. Therefore, one can conclude from Eq. (\ref{eq:ntob}) that the spectral index of the primordial tensor fluctuations for the  conventional inflation models is always negative, $n_t<0$, hence spectrum is called red-tilted~\cite{Cai:2014uka}.

\begin{figure}[H]
\centering
\subfigure[]
{\includegraphics[width=0.495\textwidth]{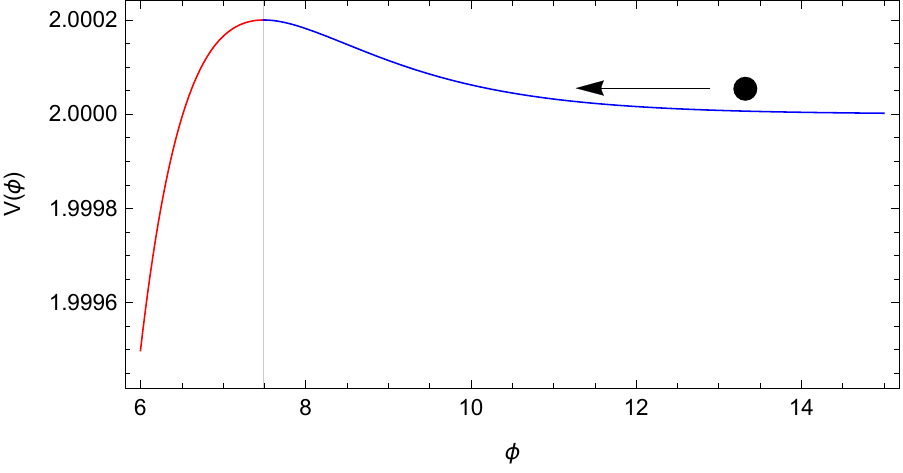}\label{fig:figpot}}
\subfigure[]
{\includegraphics[width=0.495\textwidth]{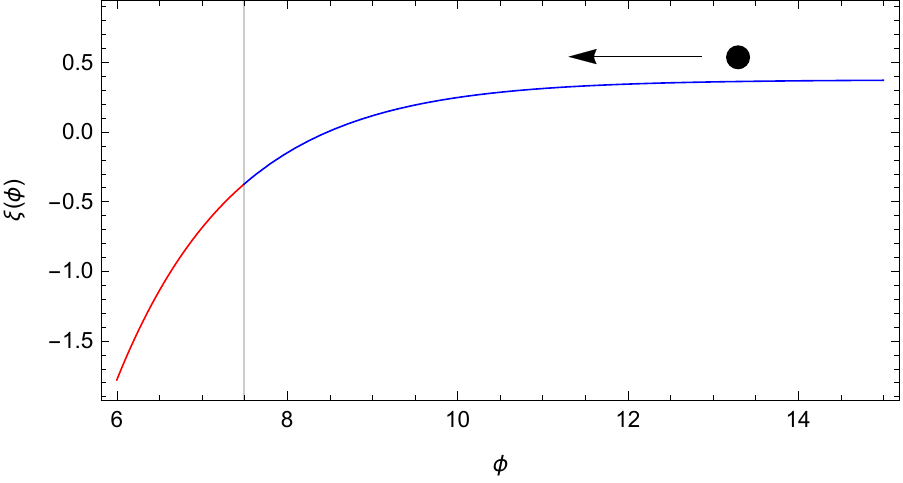}\label{fig:figxi}}
\caption{Marginalized part of the potential (left) and the Gauss-Bonnet coupling (right) shown in Figs.~\ref{fig:figp2a} and~\ref{fig:figp2b} where we set $\alpha=10^{-4}$. Vertical line corresponds to the field value, $\phi_\ast$, at which the potential takes its maximum value. At the  early stage, the Gauss-Bonnet coupling function $\xi$ makes $\phi$ climb up the potential slope. At the  late stage, $\phi$ rolls down as usual.}
\label{fig:bs}
\end{figure}

The situation is violated such that the spectrum is blue-tilted if the scalar field climbs up the potential slope, see Fig.~\ref{fig:bs}, in its early evolution before it rolls down in its late time evolution~\cite{Satoh:2010ep}.
We can see from Fig.~\ref{fig:bs} that if the initial value of the scalar field, $\phi_0$, is larger than the field value at which the potential takes its maximum, $\phi_\ast$, such that $\phi_0>\phi_{\ast}$, the scalar field needs to climb up the potential slope otherwise it simple rolls down the hill. Therefore, we argue that the blue-tilt of the spectrum for the  tensor modes would be realized when the scalar field is initially released at $\phi_0>\phi_{\ast}$.  If the  scalar field is initially released at $\phi_0<\phi_{\ast}$, the spectrum would be red-tilted. It is called the spectrum is scale invariant if $\phi_0=\phi_\ast$~\cite{Satoh:2010ep}.

On the other hand, to achieve the blue-tilted spectrum for the tensor fluctuations $n_t>0$,  $\epsilon<0$ must be satisfied from Eq.~(\ref{eq:ntob}) in our model, such that $\dot{H}>0$ is necessary from Eq.~(\ref{eq:sl_param}). Using Eq.~(\ref{eq:srpote}) together with Eq.~(\ref{eq:QofN}), one can easily obtain following condition for Gauss-Bonnet coupling function,
\bea\label{eq:condxi}
\xi_{,\phi}>-\frac{3}{4\kappa^4}\frac{V_{,\phi}}{V^2}\,,
\ena
where $V_{\phi}<0$ for the scalar field which climbs up to the potential slope. Once this conditions are satisfied, the blue-tilted spectrum of the tensor modes would be achieved in our model. By substituting Eqs.~(\ref{eq:potofphinew})--(\ref{eq:xiofphinew}) with $\beta=2$ into Eq.~(\ref{eq:condxi}), we obtain
\bea
\cosh \left(\sqrt{\frac{8\kappa^2}{q}}\phi\right) \left(\sqrt{\alpha }+\sinh \left(\sqrt{\frac{8\kappa^2}{q}}\phi\right)\right)^2>0\,,
\ena
and is clearly satisfied for all values of $\alpha>0$ and $q>0$, hence the spectrum would be blue-tilted.

\begin{figure}[H]
\centering
\subfigure[]
{\includegraphics[width=0.495\textwidth]{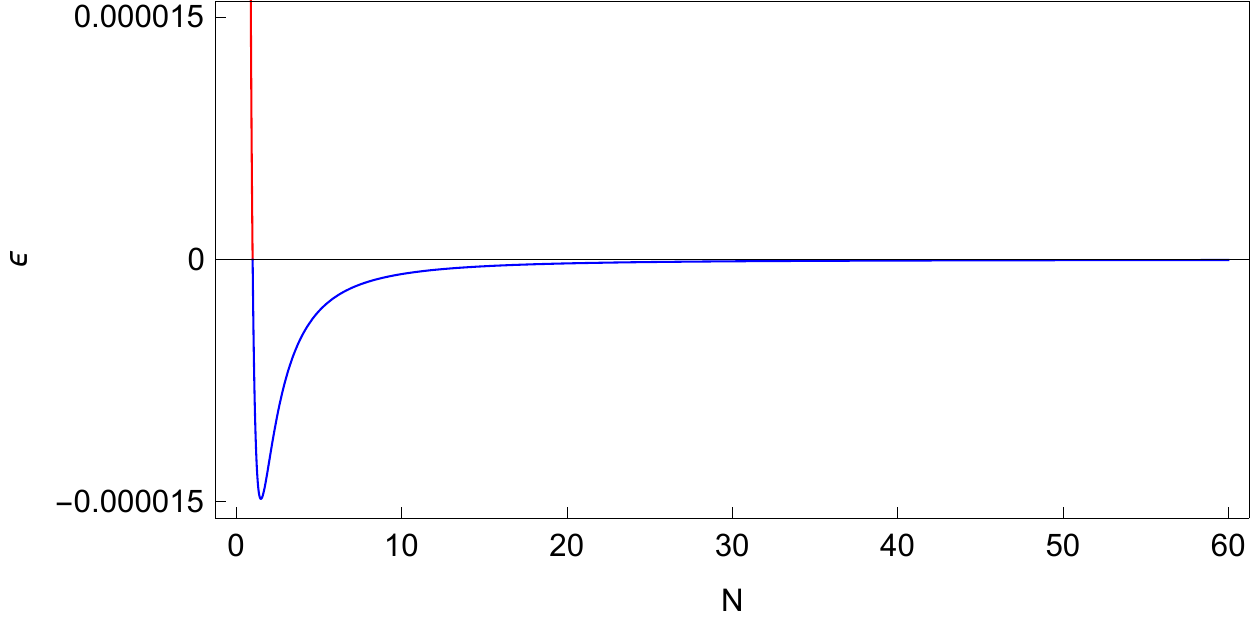}\label{fig:fig3a}}
\subfigure[]
{\includegraphics[width=0.495\textwidth]{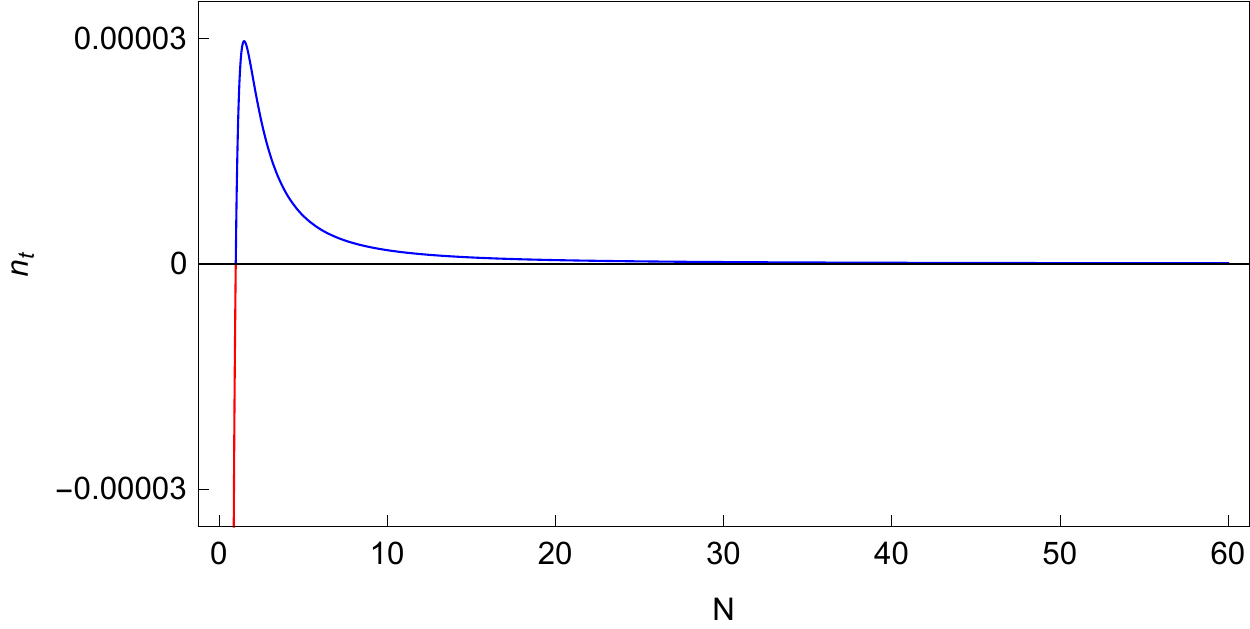}\label{fig:fig3c}}\\
\caption{$\epsilon(N)$ and $n_t(N)$ plot where we use Eqs.~(\ref{eq:potofphinew})--(\ref{eq:xiofphinew}) with $\kappa^2=1$, $c_1=1$, $c_2=0$, $\alpha=10^{-4}$, $\beta=2$ and $q=16$. At $N=1$, both $\epsilon$ and $n_t$ is zero, $\epsilon=0=n_t$.}
\label{fig:fig4}
\end{figure}

We can make the  following analysis for the blue-tilted spectrum of the primordial tensor fluctuations. As we mentioned earlier, $\epsilon$ is required to be negative for achieving the blue spectrum. In addition to this, slow-roll inflation requires the slow-roll parameters to satisfy the  slow-roll conditions in which $|\epsilon|,|\delta_1|\ll1$. Therefore, the  first slow-roll parameter takes values between $-1\ll\epsilon<0$.
After substituting Eqs.~(\ref{eq:xinew})--(\ref{eq:potofNnew}) with $p=2=\beta$ into Eq.~(\ref{srpote}), we obtain
\bea\label{eq:eps}
\epsilon&=&\frac{1}{\alpha +N}-\frac{N}{\alpha +N^2}\,,
\ena
where $\alpha>0$. We find from Eq.~(\ref{eq:eps}) the condition $-1\ll\epsilon<0$ is satisfied only for $N>1$. On the other hand, $\epsilon>0$ between $0< N<1$, hence the spectrum is red-tilted. In Fig.~\ref{fig:fig4}, we plot $\epsilon$ and $n_t$ as a function of $N$ by choosing $\alpha=10^{-4}$. The red color in Fig.~\ref{fig:fig4} between $0< N<1$, indicates positive $\epsilon$ and the red-tilted spectrum while the blue, $N>1$, color corresponds to negative $\epsilon$ and blue-tilted spectrum.

\begin{figure}[H]
\centering
\subfigure[]
{\includegraphics[width=0.495\textwidth]{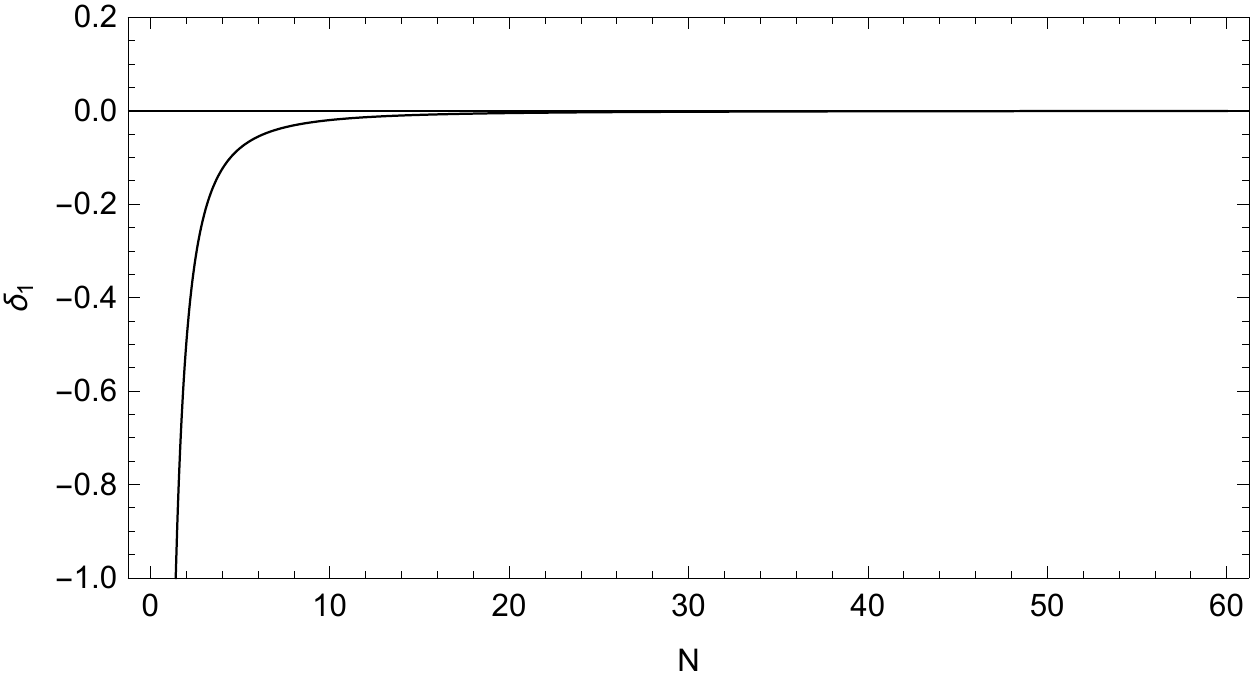}\label{fig:fig4b}}
\subfigure[]
{\includegraphics[width=0.495\textwidth]{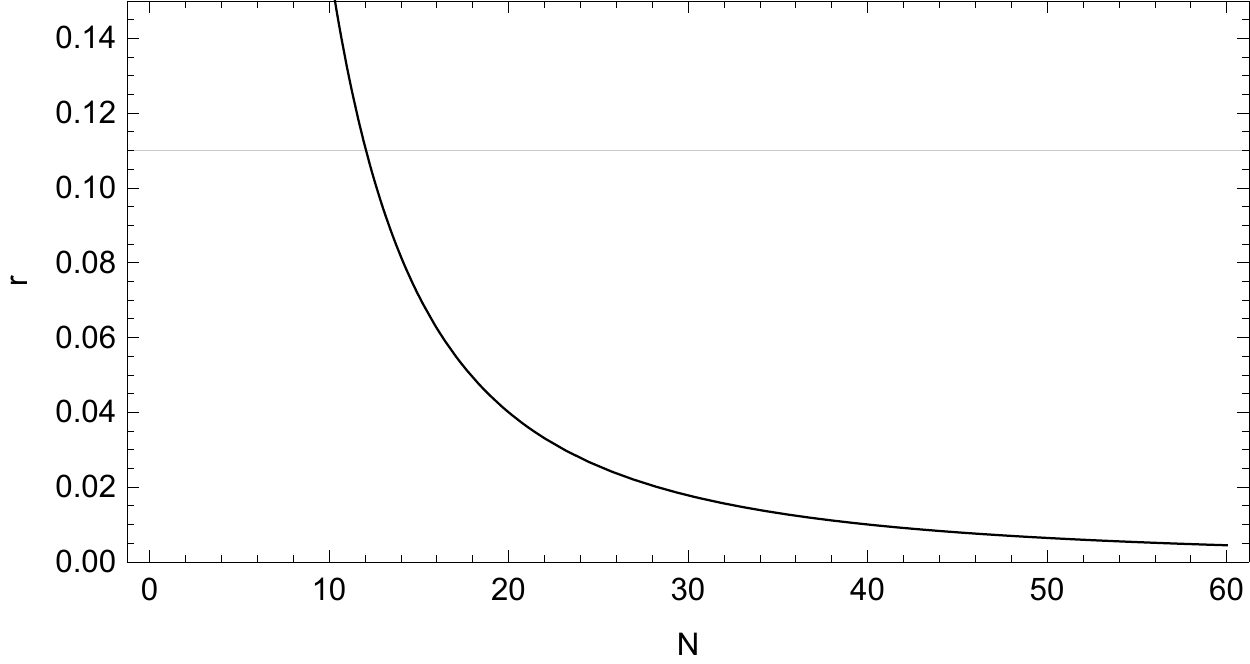}\label{fig:fig4d}}
\caption{$\delta_1(N)$ and $r(N)$ plot where we use Eq.~(\ref{eq:potofphinew}) and Eq.~(\ref{eq:xiofphinew}) with $c_1=1$, $c_2=0$, $\kappa^2=1$, $\alpha=10^{-4}$, $\beta=2$ and $q=16$. Horizontal line in Fig.~\ref{fig:fig4d} represents the current upper limit of the tensor-to-scalar ratio. } \label{fig:fig5}
\end{figure}

For the second slow-roll parameter, from Eqs.~(\ref{eq:rob})--(\ref{eq:ntob}), we find $-1\ll\delta_1<0$ because the tensor-to-scalar ratio is positive $r>0$ in our model due to our choice in Eq.~(\ref{eq:rel2}). Substituting Eqs.~(\ref{eq:xinew})--(\ref{eq:potofNnew}) with $p=2=\beta$ into Eq.~(\ref{eq:sl_d1}), we obtain
\bea
\delta_1&=&\frac{2}{\alpha +N}-\frac{16N+q}{8 \left(\alpha +N^2\right)}\,.\label{eq:del1}
\ena
Although $\delta_1$ has nothing to do with the blue spectrum for the  tensor modes, it provides a constraint on the   model parameter range for $q$. Let us search for the valid range of $\delta_1$ in which $r>0$ yields. In order the condition, $-1\ll\delta_1<0$, to be satisfied the model parameter $q$ must take values in the  following ranges:
\bea\label{eq:delcond}
\left\{
  \begin{array}{ll}
    \frac{16\alpha(1-N)}{\alpha+N}<q\leq\frac{8N^3+8\alpha N^2-8\alpha N+8\alpha^2+16\alpha}{\alpha+N} & \hbox{for $0\leq N\leq1$ }\\
\\
    0<q\leq\frac{8N^3+8\alpha N^2-8\alpha N+8\alpha^2+16\alpha}{\alpha+N} & \hbox{for $N>1$}\,,
  \end{array}
\right.
\ena
where $\alpha>0$ for both cases. We plot $\delta_1(N)$ and $r(N)$ in Fig.~\ref{fig:fig5} as an example that the model parameter $q$ which satisfies the Eq.~(\ref{eq:delcond}) gives rise to negative $\delta_1$ but positive $r$.

The most interesting and unique phenomenon for our model is that the constructed configurations of the potential and the Gauss-Bonnet coupling functions given in Eq.~(\ref{eq:potofphinew})--(\ref{eq:xiofphinew}) give rise to the blue-tilled spectral index for the  tensor modes.

\newpage
\section{Conclusion}\label{section5}

We have investigated cosmological models with a Gauss-Bonnet term to reconstruct the scalar field potential, $V(\phi)$, and the Gauss-Bonnet coupling function, $\xi(\phi)$, from the  observable quantities of $n_s$ and $r$. The main results of this work were analytically obtained in Eqs.~(\ref{eq:potofN})--(\ref{eq:xi}) where both $n_s$ and $r$ are assumed to be the  functions of $N$. We chose certain ansatz for $n_s(N)$ and $r(N)$ as seen in Eqs.~(\ref{eq:rel1})--(\ref{eq:rel2}) that are in good agreement with the  observational data~\cite{Ade:2013zuv, Ade:2015xua}. As an exercise, we considered $\gamma=1$ and $\gamma=0$ cases of Eq.~(\ref{eq:rel2}) in Section~\ref{section3}.

First we considered the model with $\gamma=1$ in Eq.~(\ref{eq:rel2}) and obtained the scalar field potential and the Gauss-Bonnet coupling functions, Eqs.~(\ref{eq:potgam1})--(\ref{eq:xigam1}). In this case, as $\alpha\rightarrow 0$ limit, our model reduces to the Einstein gravity because of the Gauss-Bonnet coupling function becomes either zero or constant such that it has no effect to the background evolution. Our result of this section is consistent with that of Ref.~\cite{Chiba:2015zpa} when $\alpha\rightarrow 0$ limit.
After this, we considered the model where $\gamma=0$ in Eq.~(\ref{eq:rel2}) and obtained the potentials and the coupling functions for $p=1$ and $p=2$, separately.

For $p=1$ case, the potential and the coupling functions have been obtained in Eqs.~(\ref{eq:potofphi})--(\ref{eq:xiofphi}). The reconstructed power-law potential shows an inverse relation to the reconstructed Gauss-Bonnet coupling function. Indeed, we obtained the  power-law potential for $\alpha\neq0$ but the minimum of the  potential shifts from zero by $\phi_{\text{min}}=-\sqrt{nq/(8\kappa^2)}$ as seen in Fig.~\ref{fig:fig1}. However, it is possible to relocate the minimum always at zero by redefining the scalar field. We also found from Eq.~(\ref{eq:xiofphi}) that the Gauss-Bonnet coupling function becomes zero if $c_2=0$ or constant if $c_2\neq0$ for every $q=4n$. In either cases, the  Gauss-Bonnet term has no effect in the background evolution and, therefore, the background evolution would determined by the Einstein gravity alone.
For $p=2$ case, we have obtained the potential and the coupling functions in Eqs.~(\ref{eq:potofphinew})--(\ref{eq:xiofphinew}). The constructed form of the  potential has a similar shape as ``T-model" in Ref.~\cite{Kallosh:2013hoa} with a small bump on the side, see Fig.~\ref{fig:figp2}. The width of the  potential is characterized by the  parameter $q$ while the height of the  bump is determined by $\alpha$ parameter. As $\alpha$ increases, the height of the bump increases and vice versa.

Another key result of our work and the most interesting feature of our model is discussed in Section~\ref{section4} where we considered the model with $\gamma=0$ and $p=2$ case. In our model, the spectrum of the primordial tensor fluctuations has been found to be blue-tilted if the newly  constructed potential and coupling functions, Eqs.~(\ref{eq:potofphinew})--(\ref{eq:xiofphinew}) with $\beta=2$ is taken to be account. This  blue-tilted power spectrum of the tensor modes, $n_t>0$, is due to the scalar field that needs to climb up its potential in the early stage of evolution hence $V_{\phi}<0$ and  $\epsilon<0$. In order to have successful inflation with enough number of $e$-folding $\sim50-60$, the scalar field in our model needs to be released at value which is larger than field value at which the potential value reached its maximum, $\phi_0>\phi_\ast$, such that it climbs up the potential slope. As a result of such climb up situation, the spectrum of the primordial tensor fluctuations would be blue-tilted.

\section*{Acknowledgements}
We appreciate APCTP for its hospitality during initiation of this work.
S.~K. was supported by the Basic Science Research Program through the NRF funded by the Ministry
of Education (No. NRF-2014R1A1A2059080).
B.~H.~L was supported by the National Research Foundation of Korea(NRF) grant funded by
the Korea government(MSIP) No. 2014R1A2A1A01002306 (ERND).

\newpage


\begin{thebibliography}{99}

\bibitem{Hinshaw:2012aka}
  G.~Hinshaw {\it et al.} [WMAP Collaboration],
  ``Nine-Year Wilkinson Microwave Anisotropy Probe (WMAP) Observations: Cosmological Parameter Results,''
  Astrophys.\ J.\ Suppl.\  {\bf 208}, 19 (2013);

  E.~Komatsu {\it et al.} [WMAP Collaboration],
  ``Seven-Year Wilkinson Microwave Anisotropy Probe (WMAP) Observations: Cosmological Interpretation,''
  Astrophys.\ J.\ Suppl.\  {\bf 192}, 18 (2011);

  E.~Komatsu {\it et al.} [WMAP Collaboration],
  ``Five-Year Wilkinson Microwave Anisotropy Probe (WMAP) Observations: Cosmological Interpretation,''
  Astrophys.\ J.\ Suppl.\  {\bf 180}, 330 (2009);
\bibitem{Ade:2013zuv}
  P.~A.~R.~Ade {\it et al.} [Planck Collaboration],
  ``Planck 2013 results. XVI. Cosmological parameters,''
  Astron.\ Astrophys.\  {\bf 571}, A16 (2014);

  P.~A.~R.~Ade {\it et al.} [Planck Collaboration],
  ``Planck 2013 results. XXII. Constraints on inflation,''
  Astron.\ Astrophys.\  {\bf 571}, A22 (2014);
\bibitem{Ade:2015xua}
  P.~A.~R.~Ade {\it et al.} [Planck Collaboration],
  ``Planck 2015 results. XIII. Cosmological parameters,''
  arXiv:1502.01589 [astro-ph.CO].

  P.~A.~R.~Ade {\it et al.} [Planck Collaboration],
  ``Planck 2015 results. XX. Constraints on inflation,''
  arXiv:1502.02114 [astro-ph.CO].

\bibitem{Guth:1980zm}
  A.~H.~Guth,
  ``The Inflationary Universe: A Possible Solution to the Horizon and Flatness Problems,''
  Phys.\ Rev.\ D {\bf 23}, 347 (1981).

  A.~Albrecht and P.~J.~Steinhardt,
  ``Cosmology for Grand Unified Theories with Radiatively Induced Symmetry Breaking,''
  Phys.\ Rev.\ Lett.\  {\bf 48}, 1220 (1982).

  A.~D.~Linde,
  ``A New Inflationary Universe Scenario: A Possible Solution of the Horizon, Flatness, Homogeneity, Isotropy and Primordial Monopole Problems,''
  Phys.\ Lett.\ B {\bf 108}, 389 (1982).

\bibitem{Martin:2013tda}
  J.~Martin, C.~Ringeval and V.~Vennin,
  ``Encyclop�dia Inflationaris,''
  Phys.\ Dark Univ.\  {\bf 5-6}, 75 (2014).
  
\bibitem{Koh:2014bka}
  S.~Koh, B.~H.~Lee, W.~Lee and G.~Tumurtushaa,
  ``Observational constraints on slow-roll inflation coupled to a Gauss-Bonnet term,''
  Phys.\ Rev.\ D {\bf 90}, no. 6, 063527 (2014);

\bibitem{Copeland:1993jj}
  E.~J.~Copeland, E.~W.~Kolb, A.~R.~Liddle and J.~E.~Lidsey,
  ``Reconstructing the inflation potential, in principle and in practice,''
  Phys.\ Rev.\ D {\bf 48}, 2529 (1993).
  

  J.~E.~Lidsey, A.~R.~Liddle, E.~W.~Kolb, E.~J.~Copeland, T.~Barreiro and M.~Abney,
  ``Reconstructing the inflation potential : An overview,''
  Rev.\ Mod.\ Phys.\  {\bf 69}, 373 (1997).

\bibitem{Chiba:2015zpa}
  T.~Chiba,
  ``Reconstructing the inflaton potential from the spectral index,''
  PTEP {\bf 2015} (2015) no.7,  073E02.

\bibitem{Cai:2014uka}
  Y.~F.~Cai, J.~O.~Gong, S.~Pi, E.~N.~Saridakis and S.~Y.~Wu,
  Nucl.\ Phys.\ B {\bf 900}, 517 (2015)


\bibitem{Kallosh:2013hoa}
  R.~Kallosh and A.~Linde,
  ``Universality Class in Conformal Inflation,''
  JCAP {\bf 1307}, 002 (2013).
  
\bibitem{Guo:2010jr}
  Z.~K.~Guo and D.~J.~Schwarz,
  Phys.\ Rev.\ D {\bf 81}, 123520 (2010)

\bibitem{Satoh:2010ep}
  M.~Satoh,
  ``Slow-roll Inflation with the Gauss-Bonnet and Chern-Simons Corrections,''
  JCAP {\bf 1011} (2010) 024.

\end{thebibliography}
\end{document}